\documentclass[prb, aps, twocolumn,showpacs,floatfix]{revtex4-1}
\usepackage{rotating}
\usepackage{color}
\usepackage{float}
\usepackage{amsmath}
\usepackage{graphicx}
\usepackage{hyperref}
\usepackage{color}
\usepackage{amssymb}
\usepackage{dcolumn}
\usepackage{bm}
\usepackage{epsfig}
\usepackage{array}
\newcommand{\be}{\begin{equation}} 
\newcommand{\ee}{\end{equation}} 
\newcommand{\bea}{\begin{eqnarray}} 
\newcommand{\eea}{\end{eqnarray}} 
\newcommand{\bwt}{\begin{widetext}}
\newcommand{\ewt}{\end{widetext}}
\newcommand{\mb}{\mathbf}
\newcommand{\mc}{\mathcal}
\newcommand{\nn}{\nonumber \\}

\newcommand{\w}{\omega}

 \newcommand{\figcdcx}
{\begin{figure}[htbp]
        \centering
        \includegraphics[angle=270,width=7.5cm]{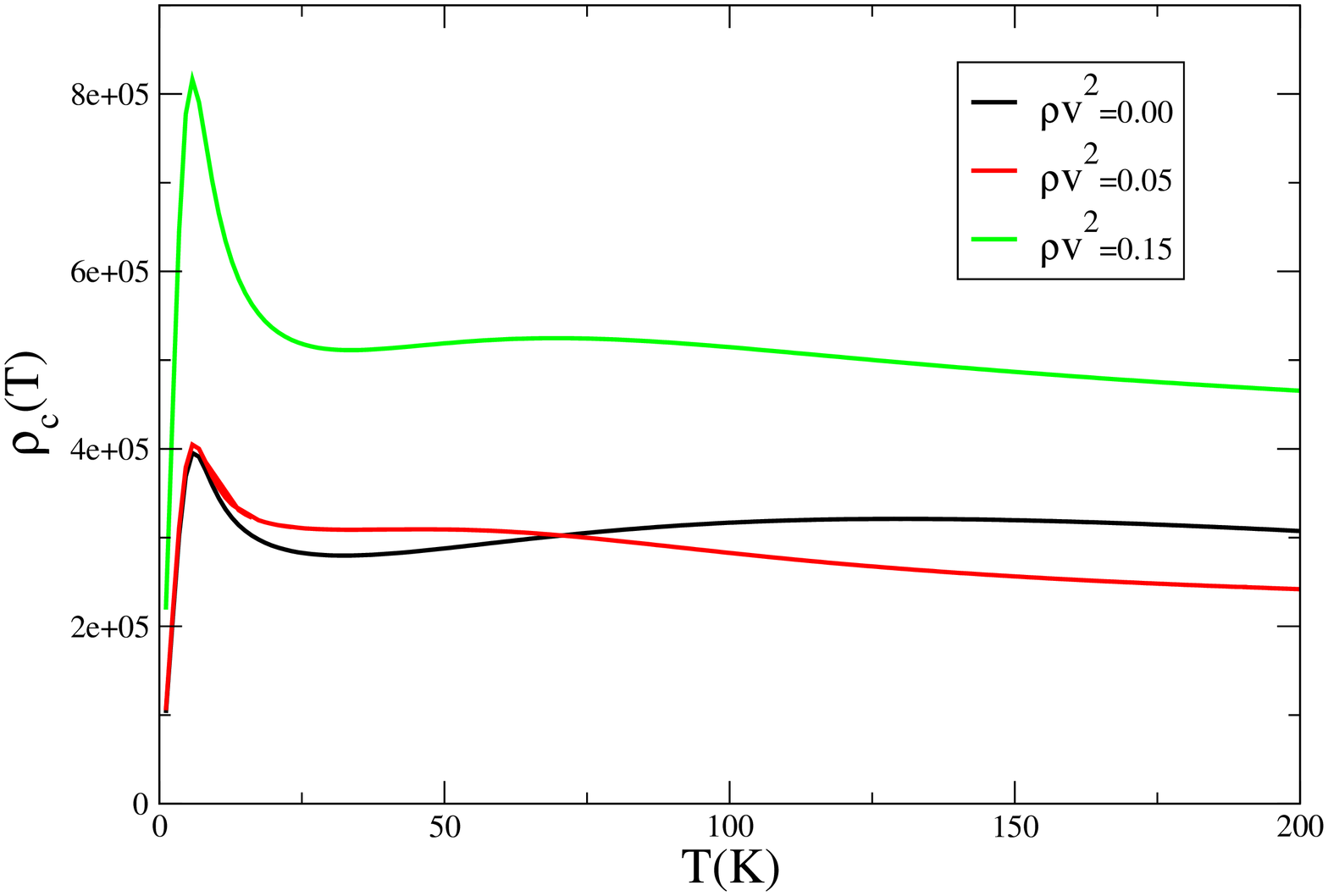}
           \caption{Inter plane or c-axis resistivity vs temperature at different impurity strengths at a fixed doping x=0.05. Here also we see upturn behavior stronger than that of the ab-plane and also gets stronger with the increase in impurity strengths.}
           \label{fig:cdcx}
	\end{figure}
} 
\newcommand{\figcdcni}
{\begin{figure}[htbp]
        \centering\includegraphics[angle=270,width=7.5cm]{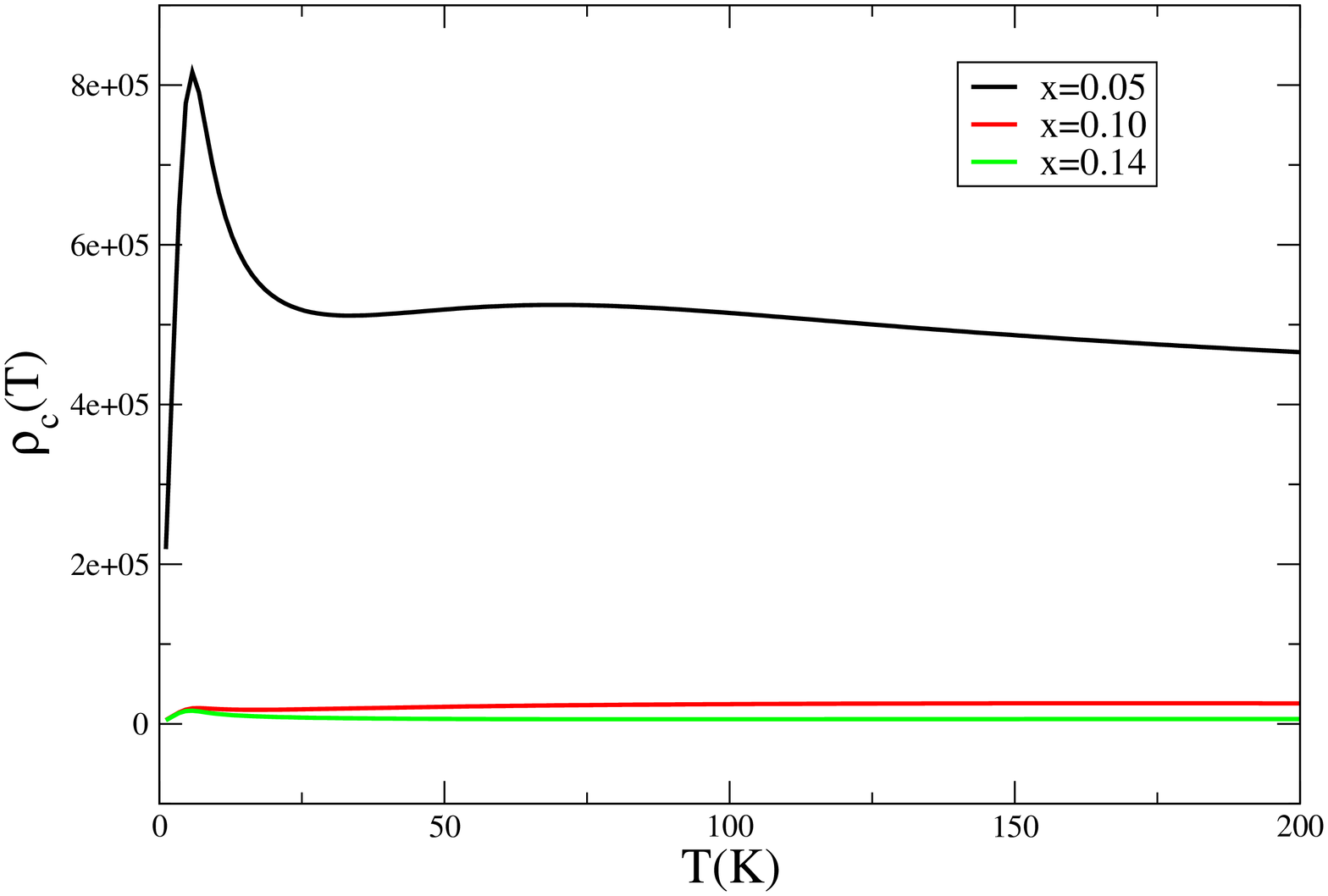}
           \caption{Inter plane or c-axis resistivity vs temperature at a fixed value of $\rho v^2=0.15$ at different hole concentrations. It also shows weaker upturn with the increase in hole concentrations as found in the case of ab-plane resistivity.}
           \label{fig:c-dc-ni}
	\end{figure}
} 
\newcommand{\figdcni}
{\begin{figure}[htbp]
        \centering
        \includegraphics[angle=270,width=7.5cm]{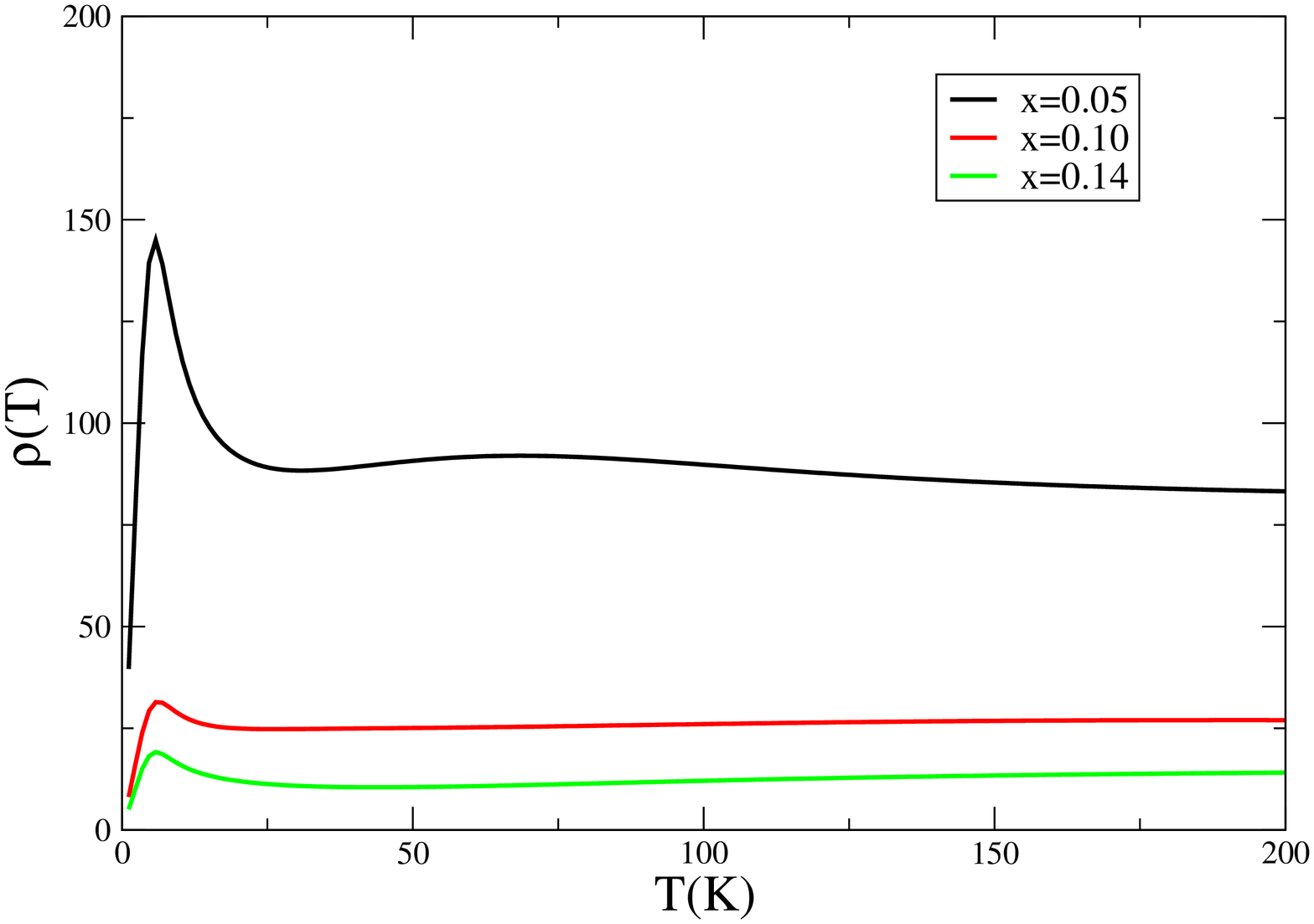}
           \caption{Resistivity in the ab-plane vs. temperature at a fixed value of $\rho v^2=0.15$ at different hole concentrations. Resistivity upturn is seen to become weaker with the increase in hole concentrations i.e. as one moves away from the pseudogap phase. }
           \label{fig:dc-ni}
	\end{figure}
} 
\newcommand{\figdcxlog}
{\begin{figure}[htbp]
        \centering
        \includegraphics[angle=270,width=7.5cm]{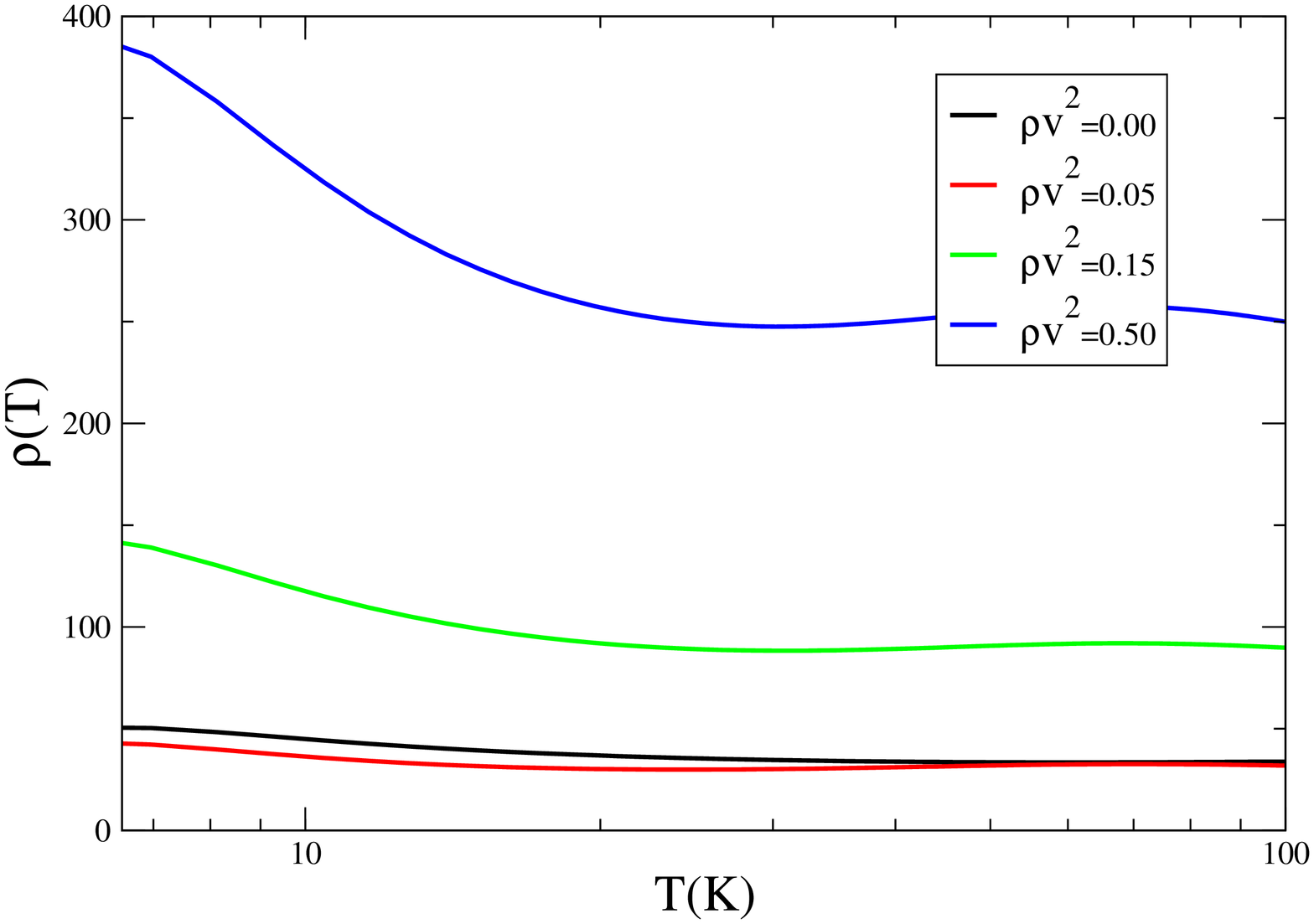}
           \caption{The ab-plane resistivity vs. temperature with $T-$axis in a logarithmic scale at different impurity strengths.
           Hole doping is fixed at x=0.05.}
           \label{fig:dcxlog}
	\end{figure}
} 
\newcommand{\figdcx}
{\begin{figure}[htbp]
        \centering
        \includegraphics[angle=270,width=7.5cm]{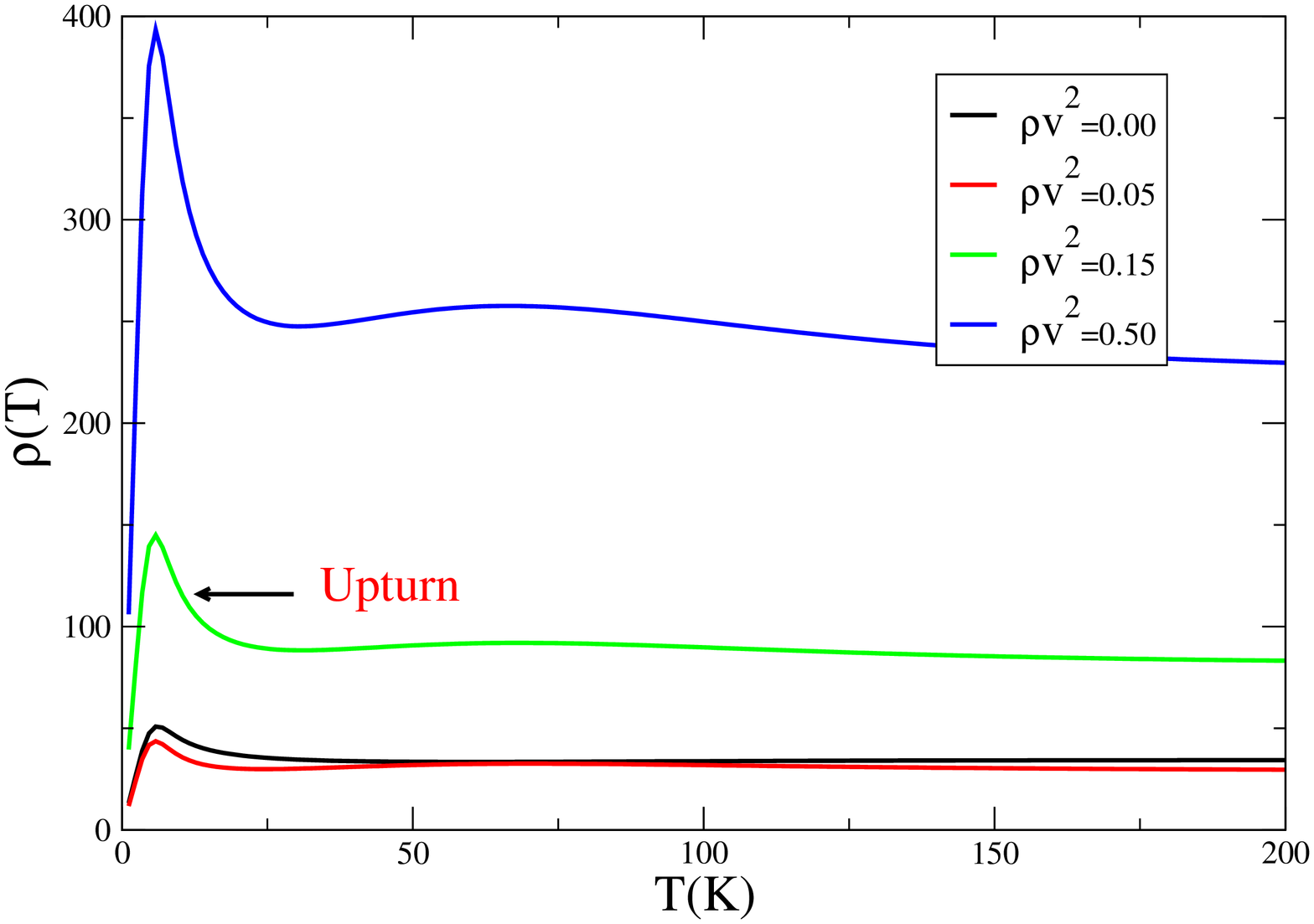}
           \caption{Resistivity in the ab-plane  vs. temperature at different impurity strengths at a fixed doping x=0.05. An upturn is seen at $\rho v^2= 0.15$ around $10$K which gets stronger in higher impurity strengths.}
           \label{fig:dcx}
	\end{figure}
}
\newcommand{\figdos}
{\begin{figure}[htbp]
        \centering
        \includegraphics[angle=270,width=8cm]{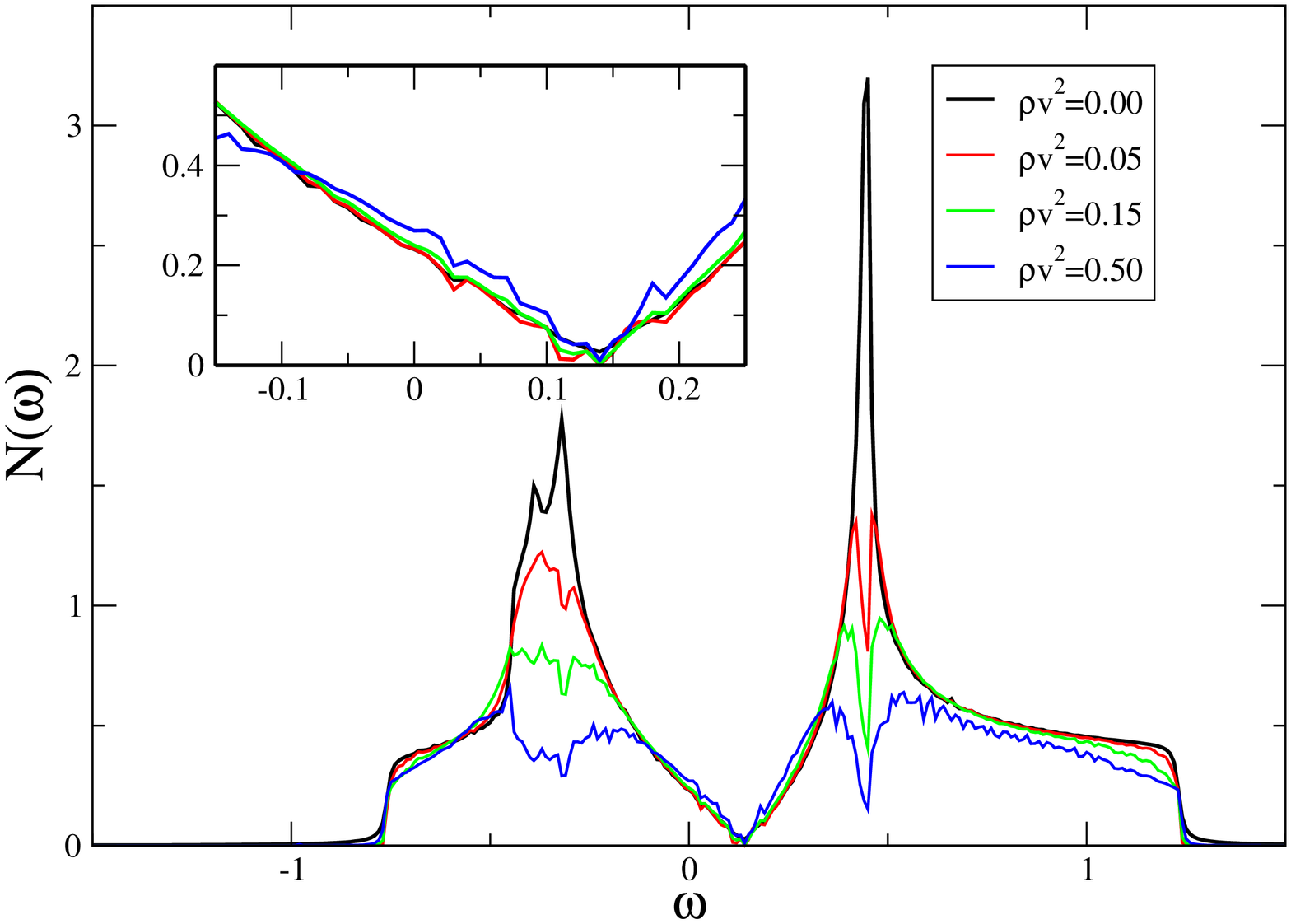}
           \caption{Un-normalised density of states of quasi-particles vs. frequency in eV at different impurity strengths. Here hole concentration x=0.05. A non-constant nature of the density of states near the Fermi surface is shown in the inset. }
           \label{fig:dos}
	\end{figure}
} \newcommand{\figimpfba}
{\begin{figure}[htbp]
        \centering
        \includegraphics[angle=0,width=7cm]{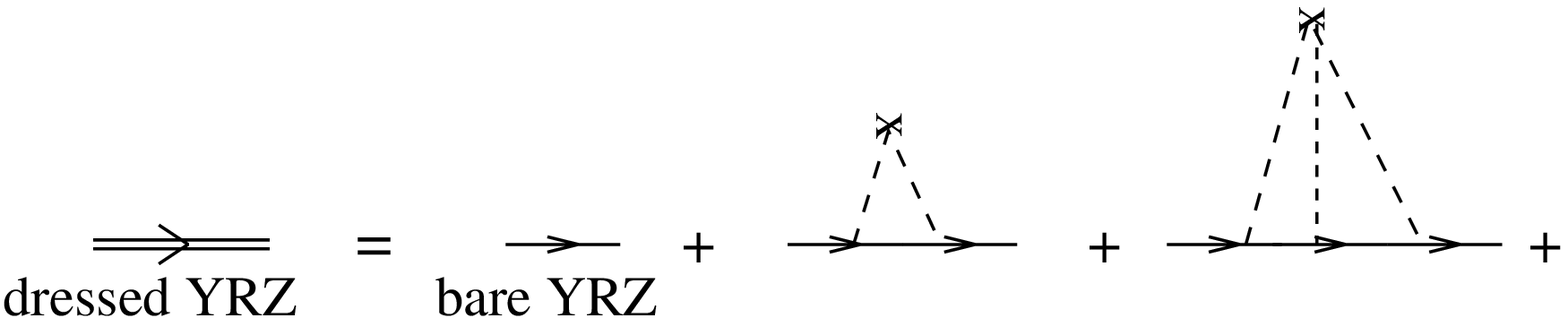}
           \caption{Diagram representing renormalized quasiparticle propagator due to impurity scattering in the Born approximation. Second and the third diagrams in the right hand side represent impurity induced self energy corrections. In the present work we consider the lowest order correction, i.e. up-to the second diagram in the right hand side. }
           \label{fig:figfba}
	\end{figure}
}

\begin{document} 
\title {Impurity induced  resistivity upturns in underdoped cuprates}
\author{Nabyendu Das}
\author{Navinder Singh}
\affiliation{Theoretical Physics Division, Physical Research Laboratory,
Ahmedabad-380009, India} 
\date{\today}

\begin{abstract}

 Impurity induced low temperature upturns in both the ab-plane and the c-axis dc-resistivities of cuprates in the pseudogap state have been observed in experiments. We provide an explanation of this phenomenon by incorporating impurity scattering of the charge carriers within a phenomenological model proposed by Yang, Rice and Zhang. The scattering between charge carriers and the impurity atom is considered within the lowest order Born approximation. Resistivity is calculated within Kubo formula using the impurity renormalized spectral functions. Using physical parameters for cuprates, we describe qualitative features of the upturn phenomena and its doping evolution that coincides with the experimental findings.  We stress that this effect is largely due to the strong electronic correlations.
\end{abstract}
\pacs{74.25.Fy, 74.72.-h }
\maketitle 

\section{Introduction}
Unlike conventional superconductors following BCS mechanism, many aspects of materials showing high temperature superconductivity are still poorly understood. Example of such materials include hole doped cuprates which show superconductivity at higher temperature. These materials show a very rich phase diagram when carrier doping and temperature vary. These materials not only lack a proper microscopic description of superconducting transition in them, but also their normal phases above the transition temperature are  poorly understood. In the over doped regime, electrons in these materials behave in a Fermi liquid manner, while a major deviation from this behavior is observed at optimal doping as well as in the underdoped regime. At low temperature, among many puzzling observations, presence of a {\it pseudogap} along with the superconducting gap in the electronic spectra is observed in this regime\cite{Timusk}. Microscopic origin of the pseudo-gap is still debatable. However 
there is hardly any doubt that many unconventional electronic behavior is associated to its presence\cite{Book}. Here we focus on understanding some anomalous charge transport properties  of these materials in the underdoped regime at low temperature. 

The effects of pseudo-gap and impurity scattering in dc-resistivities of cuprates in the underdoped regime is of our concern here. It is  well studied in the experimental literature both in presence and absence of impurities\cite{Takagi}\cite{Ando}\cite{Hanaki}. Universal linear in temperature behavior of dc-resistivity at optimal doping and its deviation both in the underdoped$(\rho\sim \log(1/T))$ and overdoped $(\rho\sim T^2)$ regimes become hallmarks for cuprates. The upturn behavior gets enhanced when cuprates are further doped with impurities and reminds us a Kondo-like mechanism as its origin. Theoretical explanation of it is not
settled yet and is highly debatable\cite{Alloul-RMP}. However, basic elements of theory of such upturn phenomena should contain two elements seriously. The first is the reconstruction of Fermi-surface and the presence of  pseudo-gap in electronic density of states  due to strong electron-electron correlations. This leads to non-Fermi liquid like behavior in the underdoped cuprates. Second is the scattering of those non-Fermi liquid 
quasi-particles with impurity atoms at low temperature. We incorporates these two important features within a semi-phenomenological theory in the present study.

Here we limit ourselves to the situations where cuprates are doped with non-magnetic impurities only. Experimentally, cuprates are further doped with non-magnetic Zn ions which replaces few Cu ions from Cu-O plane and provide a scattering center for charge carriers. Doped Zn ions reduces superconducting correlations to certain extent. Also a magnetic field of suitable strengths (up-to $\sim 50-60T$) are used to exclude the contributions due to superconducting correlations. This sets the platform for studying normal state properties of cuprates which often shows non-Fermi liquid behavior, particularly in the under-doped regime. It is also to be mentioned that, effects similar to the Zn doping can be achieved by another method such as electron irradiation with suitable fluence\cite{Alloul-RMP}. In the later case one creates point defects in the Cu-O plane and the gross features of the resistivity behavior in this case is almost the same as that of Zn substituted one. In such an experimental condition, depending on impurity concentrations, an impurity induced supra-linear in $T\,(\rho\sim T^\nu, \, \, \nu<1)$ to $\log(1/T)$ upturn in the resistivity is observed\cite{Hussey}\cite{Boebinger}.  The later phenomenon hints at localization onset and are often attributed to strong electron-electron correlations, spin fluctuation effects, impurity induced electronic inhomogeneity\cite{Segawa}\cite{Alloul1}\cite{Alloul2}.
Since a detailed microscopic theory for underdopped cuprates still lacking, logical direction in understanding this material is to use a phenomenological
theory based on experimental inputs. Recently Yang, Rice and Zang\cite{YRZ-PRB}\cite{YRZ-RPP} proposed a phenomenological model which can be used to understand properties of electronic behavior in the underdoped regime.  This theory is quite successful in understanding many thermodynamic properties and ARPES findings\cite{YRZ-RPP} and also transport properties like optical conductivity\cite{Illes}\cite{Navinder} etc. in  the recent past. We  adopt it as a model non-Fermi liquid description suitable to study low temperature behavior of normal state resistivity in cuprates in presence of a pseudo-gap. For the rest of our discussion we will call it ``YRZ'' formalism and in the later section we present a brief review of it.
In this work, we show that, without adopting additional inputs one can observe resistivity upturns within YRZ formalism. Also the present study reveals the importance of pseudo-gap to this phenomena. 

This paper is organized as follows. In section \ref{sec:theory} we first review the YRZ formalism as proposed by  Yang, Rice and Zang\cite{YRZ-PRB}\cite{YRZ-RPP}. and then derive an expression for the dc-resistivity using Kubo formalism and Considering the scattering of YRZ-quasiparticles with non-magnetic impurities. Then in section \ref{sec:results}, we numerically evaluate the expression for the dc-resistivity and discuss the results. Finally, in section \ref{sec:summary} we summarize our results and conclude.
\section{YRZ-quasiparticles and Impurity scattering}
\label{sec:theory}
In the presence of a pseudogap, YRZ formalism is defined with the following ansatz for the coherent part of the electronic self energy based on a resonating valence bond (RVB) spin liquid\cite{YRZ-PRB}\cite{YRZ-RPP}.
\be
\Sigma^{YRZ}(\mb{k}, \omega, x)=\frac{\Delta^2_{pg}}{\omega+\epsilon^0_{\mb{k}}}
\ee 
Thus the quasiparticle propagator or Green’s function in this formalism takes the form,
\be 
\mc{G}(\mb{k}, \omega, x)=\frac{g_t(x)}{\omega-\epsilon_{\mb{k}}-\frac{\Delta^2_{pg}}{\omega+\epsilon^0_{\mb{k}}}} .
\label{eq:eqyrz0}
\ee
The interaction renormalized band-structure dispersion is given as $\epsilon_{\mb{k}} = - 2t(x) (\cos(k_xa)+\cos(k_ya))-4t'\cos(k_xa) \cos(k_ya)-2t''(\cos(2k_xa)+\cos(2k_ya))-\mu_p(x)$.
Here $a$ is the unit cell dimension in the Cu-O plane. Electronic energy band is defined by the renormalized parameters,
$t(x) = g_t(x)t_0 + 3g_s(x)J\chi/8,\,\, t'(x)= g_t(x)t_0'\,{\rm and}\,\, t''(x)=g_t(x)t_0''$. 
The Gutzwiller factors which constrain electronic motion in an interacting environment are given by,
 $ g_t(x)=\frac{2x}{1+x}$ and $ g_s(x)=\frac{4}{(1+x)^2}$.
 Here, $t_0, \, t_0',\,{\rm and}\, t_0''$ are the bare band hopping parameters with
$t_0'=−0.3t_0 \,{\rm and}\, t_0''=0.2t_0$.
 $J$ is the magnetic energy of the $t$-$J$ model taken to be $J=t_0/3,\,\chi=0.338$ is the spin susceptibility.
 $\epsilon^0_{\mb{k}}=-2t(x)\\ (\cos(k_xa)+\cos(k_ya))$ is a second energy dispersion which gives the antiferromagnetic
Brillouin zone boundary for  $\epsilon^0_{\mb{k}}=0$, which is also referred
to as the umklapp surface.
 The shift in chemical potential $\mu_p(x)$ is to be determined to
get the correct hole doping $x$ based on the Luttinger sum
rule. Self-energy contains a phenomenological input for pseudo-gap energy scale and is given as, $\Delta_{pg}(x)=\Delta^0_{pg}(x)(\cos(k_xa) -\cos(k_ya))/2$, where $\Delta^0_{pg}(x)=3t_0(0.2 - x)$. In this description $x=0.2$ is the optimal doping and we restrict to the regime $x<0.2$. In the normal state, the quasiparticle propagator can be written in a form,
 \bea 
 \mc{G}(\mb{k}, \omega, x)= \sum_{\alpha=\pm 1}\frac{g_t(x)W^\alpha_{\mb{k}}(x)}{\omega-E^\alpha_{\mb{k}}(x)}\cdot
 \label{eq:eqyrz}
 \eea Above form describes YRZ quasiparticles by the energy dispersions,
 \bea 
 E^\alpha_{\mb{k}}(x)=\frac{\xi^-_{\mb{k}}}{2}+\alpha \sqrt{\left(\frac{\xi^+_{\mb{k}}}{2}\right)^2+\Delta^2_{pg}(x)},
 \eea 
 and the Luttinger weights are given as,
 \bea 
 W^\alpha_{\mb{k}}(x)=\frac{1}{2}\left( 1+\alpha\frac{1}{2} \frac{\xi^+_{\mb{k}}}{\sqrt{\left(\frac{\xi^+_{\mb{k}}}{2}\right)^2+\Delta^2_{pg}(x)}}\right).
 \eea  
 Here $\alpha=\pm1$ and $\xi^\pm_{\mb{k}}=\epsilon_{\mb{k}}\pm\epsilon^0_{\mb{k}}$.
We consider  scattering of YRZ quasi-particles with non-magnetic impurities in the first Born Approximation.  They provide scattering centers for  quasi-particles and changes their lifetime via impurity contribution to the self-energy. Diagrammatic representation of the self energy correction due to impurity scattering is shown in Fig.\ref{fig:figfba}.
 \figimpfba
Here arrowed parallel lines in the left hand side represents impurity renormalized propagator. The first diagram in the right hand side is the bare YRZ propagator as given by Eq.\ref{eq:eqyrz} and the third term represents self energy corrections to the carrier propagator due to its interaction with impurities at the lowest order. The later is given as\cite{Doniach},
 \bea 
 \Sigma_{imp}(\mb{k},i\w_n)&=&N_i\sum_\mb{k'} \left|v(\mb{k}-\mb{k'})\right|^2 \mc{G}(\mb{k'},i\w_n)\nn 
                                 &\approx& \rho v^2\frac{1}{\mc{V}}\sum_\mb{k'} \mc{G}(\mb{k'},i\w_n)\nn
                                 &=&\rho v^2 \frac{1}{\mc{V}}\sum_{\alpha=\pm}\sum_\mb{k'}\frac{g_t(x) W^\alpha_{k'}}{i\w_n-E^\alpha_{k'}}\cdot 
 \label{eq:simp-dos0}                                
 \eea 
  Here impurities potential is assumed to be highly localized i.e. given by $v(\mb{r})=v\delta(r)$. This is justified when doped ions sit in the Cu-O plane and charge carriers screen the Coulomb potential. Impurity density $N_i/\mc{V}$ is given by $\rho$ and $\mc{V}$ is the system volume. After performing an analytical continuation we get the frequency dependent impurity self energy as,
 \bea 
 \Sigma_{imp}(\w)&=& \sum_{\alpha=\pm}\sum_\mb{k'}\frac{\tilde{W}^\alpha_{k'}}{\w-E^\alpha_{k'}+i\eta} \nn
 &=& \sum_{\alpha=\pm}\sum_\mb{k'}\frac{\tilde{W}^\alpha_{k'}}{\w-E^\alpha_{k'}} -i\pi \sum_{\alpha=\pm}\sum_\mb{k'}\tilde{W}^\alpha_{k'}\delta(\w-E^\alpha_{k'})\cdot                     
 \eea 
Where $\tilde{W}^\alpha_{k}=\rho v^2 g_t(x)W^\alpha_{k}/\mc{V}$. This will be different from that of non-interacting case. In case of normal metal, and for constant electronic density of states, the real part of the Born self energy vanishes, while imaginary part of it becomes constant\cite{Sadovskii}. That is not necessarily the case here.  We see from the above equation that,
 \bea 
 Im\Sigma_{imp}(\w)&=& - \pi\sum_{\alpha=\pm}\sum_\mb{q}\tilde{W}^\alpha_{q}\delta(\w-E^\alpha_{q}) \nn
                           &=&- \rho v^2 N^{YRZ}(\w)\cdot
\label{eq:simp-dos}
 \eea 
 Here $N^{YRZ}(\w)$ is the un-normalized density of states for YRZ quasiparticles. 
In previous works\cite{Navinder}\cite{Pankaj}, where optical conductivity of cuprates in the pseudogap state was studied, $\gamma_0=1/\tau_0$ was set to be $0.01t_0$ to regularize the delta-function in the expression of the quasiparticle density of states\cite{Illes}\cite{Ashby}. Considering impurity scattering within lowest order Born approximation, we define impurity renormalized spectral function of quasi-particles as,
\bwt
 \be
 \mc{A}(\mb{k},\w)=-\frac{1}{\pi}g_t(x)\frac{Im\Sigma_{imp}(\w)}{\left(\w-\epsilon_{\mb{k}}-\Sigma^{YRZ}(\mb{k},\w)-Re\Sigma_{imp}(\w)\right)^2+(Im\Sigma_{imp}(\w))^2} \cdots
 \ee
 \ewt
 The spectral function above includes the incoherent component to the bare YRZ spectral function. The incoherent part comes from scattering of the  quasi-particles with localized impurity atoms.  Now we use Kubo formula in the linear response regime to numerically calculate the conductivity. 
 For a $N\times N$ lattice, the expression for conductivity at a finite frequency and temperature is given as follows.
 \bwt
 \be
 \sigma(\w,T)= \frac{4\pi e^2 \hbar}{N^2ca^2}\sum_{\mb{k}=\left(-\frac{\pi}{a},-\frac{\pi}{a}\right)}^{\left(\frac{\pi}{a},\frac{\pi}{a}\right)}
 v^2_{0,x}(\mb{k})\int_{-\infty}^{\infty}d\w'\frac{f(\w')-f{(\w+\w')}}{\w}\mc{A}(\mb{k},\w')\mc{A}(\mb{k},\w+\w')
 \ee
 \ewt
 We are concerned about dc-resistivity behavior, we consider the $\w\rightarrow 0$ limit of the above expression. Thus at finite temperature, dc conductivity or zero-frequency conductivity takes the following form,
 \bwt
 \bea 
 \sigma_{dc}(T)= \frac{4\pi e^2 \hbar}{N^2ca^2}\sum_{\mb{k}=\left(-\frac{\pi}{a},-\frac{\pi}{a}\right)}^{\left(\frac{\pi}{a},\frac{\pi}{a}\right)}
 v^2_{0,x}(\mb{k})\int_{-\infty}^{\infty}d\w'\left(-f'(\w')\right)\mc{A}(\mb{k},\w')^2\cdot
 \label{eq:dc}
\eea 
\ewt
 DC-resistivity $\rho$ is calculated by taking the inverse of $\sigma_{dc}(T)$. Here the frequency derivative of the Fermi-Dirac distribution function is given as,
 \bea 
 f'(\w) &=&\frac{\partial}{\partial \w}\left(\frac{1}{e^{\w/T}+1}\right)= \frac{\partial}{\partial \w}\left(\frac{1}{2}-\frac{1}{2}\tanh\left(\frac{\w}{2T}\right)\right)\nn
 &=& -\frac{1}{4T} \rm{sech}^2\left(\frac{\w}{2T}\right)\cdot
 \label{eq:Fermi}
 \eea 
 This function peaks near $\w=0$ and has width $2T(k_B=1)$. Thus for numerical calculations, we can fix our $\w$-cutoff in calculating dc-conductivity from Eq.\ref{eq:dc} as few $T$.\\ 
\noindent
{\it C-axis conductivity.}
In Cuprates, charge carriers interact in the Cu-O or ab-plane and studies related to these materials are limited to two dimensional model. However in a real material, such Cu-O planes stack along c-axis and carriers can tunnel between two neighboring planes through tunneling mechanism. Conduction along c-axis is also studied extensively\cite{Book}. In the case of non-magnetic impurity doped samples,
an upturn in the low temperature regime is also observed\cite{Ando}. We revisit the same here within our formalism. To calculate conductivity along c-axis, we follow Chakravarty et al. \cite{sudbo} and introduce tunneling matrix element between adjacent Cu-O layers which will replace $v_x$ in the earlier expressions for conductivity in Eq.\ref{eq:dc}  with,
\bea 
 t_{\perp}(\mb{k})c=\frac{d t_{\perp}}{4}\left(\cos(ak_x)-\cos(ak_y)\right)^2.
\eea 
Here $t_{\perp}$ is the tunneling matrix element for quasiparticles in cuprates along the c-axis. It is usually in the range $0.1-0.15$eV. The parameters $c$ and $a$ are the lattice dimensions along the c axis and a or b-axis. With these descriptions, we conclude our formalism.

\section{Results}
\label{sec:results}
In this section we discuss the findings of our formalism. We show plots of various quantities for $100\times 100$ lattice and for $t_0=1$ eV. Other parameters are taken from previous works by one of the present authors\cite{Navinder}\cite{Pankaj}. Coefficient of Eq.\ref{eq:dc} is set such that $\rho_{ab}(T)$ is given in the $\mu\Omega \,{\rm .cm}$ unit.  Before presenting the results, we estimate the validity of our approximation. We consider the lowest order correction to the quasiparticle self energy that gives a frequency dependent contribution and limit ourselves to the 2nd diagram in the right hand side of the Fig.\ref{fig:figfba}.  Next order correction, as shown in the  diagram next to it, is given as  $\rho v^3\left( N^{YRZ}\right)^2 $ while the  previous diagram gives a $\rho v^2\left( N^{YRZ}\right)$ contribution\cite{Doniach}. Thus our perturbative approach depends on the condition that the ratio of this two successive contributions, i.e.  $v\times N^{YRZ}<<1 $. In our system we see that near the Fermi level, $N^{YRZ}\sim 0.1$ per eV and if we assume the energy scale 
of the coupling between impurities and quasiparticles $v\sim 100$ meV, then  $v\times N^{YRZ}\sim 0.01<<1$ and the condition is well satisfied.
 Within this approximation, impurity induced correction  to the self energy is determined by a parameter $\rho v^2$ than $v$ alone. We consider it as a free parameter and vary it within certain range assuming that earlier conditions are satisfied.

In Fig.\ref{fig:dos} we show plots of momentum integrated un-normalized density of states at $x=0.05$ as a function of energy for various impurity strengths. It is observed that density of states has a valley near $\w\sim \mu$ and two peaks are separated by twice the pseudo-gap energy scale, i.e.  $2\Delta^0_{pg}$. However since the pseudogap is not constant for all momentum states and contains nodes and anti-nodes in it's structure, it affects the density of states profile within the range $(- \Delta^0_{pg}, \Delta^0_{pg})$. Under its influence density of states near the Fermi surface never vanishes, rather becomes energy dependent. This is one aspect where the pseudogap physics becomes different from that of the simple metals.  The plot for $\rho v^2=0.00$ corresponds to the bare quasiparticle case where a regularization in the Dirac delta function was made by introducing an artificial width $\gamma_0=0.01t_0$.  In this figure we also plot the density of states for various impurity strengths and confirm that coherence in the quasiparticle density of states survives within our choices of impurity strengths. In the inset of the same figure we show the behavior of the density of states near the Fermi surface within the energy scale of our interest. It clear that the unlike the case of metal, density of states is strongly energy dependent in this regime as expected in the pseudogap regime.
 
\figdos

From our experiences in calculating transport coefficients in metals, it is tempting to relate the density of states to the scattering rate of the quasiparticles or the resistivity. However from Eq.\ref{eq:dc} and Eq.\ref{eq:Fermi} we see that it is not exactly the electronic density of states near the Fermi level$(\omega=0)$ within the energy scale $\sim T$ ($\sim$ few meV) determines the scattering rate of the quasiparticles and hence the resistivity. A quantity which we can name transport density of states or tDOS$=\sum_\mb{k} v_k^2 \mc{A}(\mb{k}, \w)^2$, as opposed to the density of states obtained from the spectroscopic experiments and is given by $\sum_\mb{k}  \mc{A}(\mb{k}, \w)$ determines the scattering rates and hence resistivity. In Fig.\ref{fig:dos} we see that the density of states near the Fermi level within an energy interval of few meV is strongly energy dependent. Also, with increasing impurity strength, the peaks in the density of states profile decrease and the spectral weights are transfered towards the Fermi level. At stronger impurity strength, the curvature of the density of states profile near $\w=0$ changes. Since $\mc{A}(\mb{k},\w)$ is positive for all $\mb{k}$, all these features, though in a qualitatively different way, should manifest in the tDOS also. Now for qualitative understanding, we can assume $tDOS\sim (a-b\,\rm{sgn}(\w)|\w|^\alpha)$, where $a,b>0$ and $\alpha$ are  constants.
Using Eq.\ref{eq:dc}, we see $\sigma\sim \int_{-\Lambda}^\Lambda d\w (a-b\,\rm{sgn}(\w)|\w|^\alpha)\, \rm{sech}^2(\w/T)$. The thermal factor is positive for all the values of $\w$ and behaves like a Gaussian whose width is given by $T$. Now scaling $\w$ with temperature, we get $\frac{1}{\rho}=\sigma\sim \int_{-\infty}^\infty dx (aT-bT^{1+\alpha} \rm{sgn}(x)|x|^\alpha)\, \rm{sech}^2(x)=AT-BT^{1+\alpha}$ in the limit $T<<\Lambda$. Here $\alpha$ is determines the curvature of the tDOS profile and along with $B$, determines the deviation from the uniformity. A more appropriate fitting of tDOS should contain other powers of $\w$ and they will also contribute to the resistivity. It is also nice to comment here that, if for some interactions tDOS of quasi-particles fits with $\alpha=-2$, resistivity becomes linear in $T$. Here all the constants have dependence on the pseudogap, hence doping as well as the impurity scattering. In our numerical calculations we see that the competition between the such terms as observed in the above expression, give rise to the upturn. The dependence on the velocity $v_x$ in tDOS is also reflected in qualitative differences between resistivity profiles in $ab$-plane and along $c$-axis, as is shown in our results.
 \figdcx
 \figcdcx
 \figdcxlog
We present the plots for resistivity with temperature at different impurity strengths and doping and focus on the effects of pseudo-gap phase which is most prominent at low doping. We first consider $x=0.05$, and look for resistivity vs. temperature behavior at different impurity strengths. In Fig.\ref{fig:dcx} and Fig.\ref{fig:cdcx}, we see weak upturn for both $\rho_{ab}$ and $\rho_c$ at low temperature, which increases with increasing impurity strength. To study the appearance of a $\log(1/T)$ behavior in resistivity associated with this upturn, we re-plot the Fig. \ref{fig:dcx} with temperature axis in a logarithmic scale and is shown in Fig.\ref{fig:dcxlog}. We see that at higher impurity strength, resistivity profile in the upturn regime  moves toward a $\log(1/T)$ one.  This is in accord with experimental findings\cite{Segawa}.  
 \figdcni
 \figcdcni 
 Next, to study the effects of carrier concentration or pseudo-gap in the upturn formation we plot again resistivity vs temperature at different doping with fixed impurity strength. We choose $\rho v^2=0.15$, as an upturn is visible at this impurity strength. The ab-plane and the c-axis resistivities are plotted for various doping in Fig.\ref{fig:dc-ni} and Fig. \ref{fig:c-dc-ni} respectively. We see that, in both cases, the upturns get weaker with increasing hole concentrations, i.e. as one moves away from the pseudo-gap phase. 
\section{Summary and conclusions}
\label{sec:summary}
The main observations from this work are as follows. Unlike the case of metals, impurity contribution to the quasi-particle self energy in the pseudo-gap phase of cuprates is  frequency dependent and is determined by the density of states profile. Replacing  the constant width of the single particle states by this frequency dependent self-energy, can lead to an upturn in the temperature variation of the resistivity at low temperature which is often considered as precursor to a metal-insulator transition. Also with higher impurity strengths, i.e. higher values of $\rho v^2$, we see a proximity towards $\log(1/T)$ like behavior in resistivity. Variation of the upturn with carrier concentration is  qualitatively in accord with experiments. Now we compare our formalism  with previously proposed theories.  We observe,  emphasis on either or a combination of the 1) Kondo like scenarios, 2) weak localization corrections, 3) impurity induced inhomogeneous electronic distribution picture and 4) pseudogap and strong electron-electron correlations in all the proposed theories depending on the regime of their interest in the cuprate phase diagram\cite{Kopp}\cite{Kontani}\cite{Chen}. For example authors in Ref.\cite{Kontani}\cite{Chen} focus on optimally and slightly underdoped cuprates and develop their theory and studies the disorder contribution to the upturn phenomenon in a metalic Fermi liquid background. In their studies they limit electronic correlations within weak interaction regime while disorder contribution is calculated within a numerically exact way. They find importance of impurity induced magnetic droplets in the formation of resistivity 
upturn. Though such a picture may be valid near optimal doping, but certainly not in the deep underdoped regime where strong electron-electron correlations is of utmost interest. We focus on the later regime. Since there is no well accepted microscopic description valid in this regime, we adopt a simpler phenomenological picture, namely YRZ-theory to incorporate non-perturbative aspects of electron-electron correlations. Thus we show that even if we limit ourselves to the perturbative Born approximation to consider the elastic scattering between strongly correlated electrons, the upturn behavior in resistivity can appear at low temperature. This is certainly a notable observation which puts more weight to the electronic correlations than the disorder effects. From our analysis, we  conclude that the presence of the pseudo-gap is enhancing the upturn phenomena which is also observed in the previous theoretical proposal\cite{Chen}. Observed $\log(1/T)$ like behavior of resistivity at higher impurity concentrations is also a prominent experimental feature in these materials and its origin is highly debatable\cite{Segawa}. However, it  needs to be mentioned that in this work, we limit 
ourselves to {\it isotropic, zero range, non-magnetic} impurities. More realistic impurities providing momentum-dependent impurity potential\cite{Zhu} and impurities with net magnetic moment, e.g. $Ni$ doping\cite{Tanabe} are surely worth studying. These are left for future studies.

\end{document}